\documentclass[runningheads]{llncs}

\usepackage{graphicx}
\usepackage{amsfonts}
\usepackage{amsmath}
\usepackage{amssymb}
\usepackage{enumerate}
\usepackage{bm}
\usepackage{color}
\usepackage{wrapfig}

\usepackage{hyperref}
\begin{document}

\title{Algorithmically generating new algebraic \\ features of polynomial systems \\ for machine learning
%\thanks{The authors are supported by EPSRC Project EP/R019622/1: \emph{Embedding Machine Learning within Quantifier Elimination Procedures}.}
}

\titlerunning{Generating new features of polynomial systems for machine learning}
% If the paper title is too long for the running head, you can set
% an abbreviated paper title here
%

\author{Dorian Florescu \and Matthew England}
\authorrunning{D. Florescu and M. England}
% First names are abbreviated in the running head.
% If there are more than two authors, 'et al.' is used.

\institute{Faculty of Engineering, Environment and Computing, \\Coventry University, Coventry, CV1 5FB, UK
\email{\\ \{Dorian.Florescu, Matthew.England\}@coventry.ac.uk}}

\maketitle              % typeset the header of the contribution

\begin{abstract}
There are a variety of choices to be made in both computer algebra systems (CASs) and satisfiability modulo theory (SMT) solvers which can impact performance without affecting mathematical correctness.  Such choices are candidates for machine learning (ML) approaches, however, there are difficulties in applying standard ML techniques, such as the efficient identification of ML features from input data which is typically a polynomial system.  
Our focus is selecting the variable ordering for cylindrical algebraic decomposition (CAD), an important algorithm implemented in several CASs, and now also SMT-solvers.  We created a framework to describe all the previously identified ML features for the problem and then enumerated all options in this framework to automatically generation many more features.  We validate the usefulness of these with an experiment which shows that an ML choice for CAD variable ordering is superior to those made by human created heuristics, and further improved with these additional features.  We expect that this technique of feature generation could be useful for other choices related to CAD, or even choices for other algorithms with polynomial systems for input.

%<250 words (currently 228)


\end{abstract}

\section{Introduction}
\label{SEC:Intro}

Machine Learning (ML), that is statistical techniques to give computer systems the ability to \emph{learn} rules from data, is a topic that has found great success in a diverse range of fields over recent years.
%, such as well publicised applications in healthcare \cite{CHHWW17} and text analysis \cite{YSZ17}.  Software engineering itself is affected, in particular the testing and security analysis stages \cite{GS17}.  
%
ML is most attractive when the underlying functional relationship to be modelled is complex or not well understood.  Hence ML has yet to make a large impact in the fields which form \textsf{SC}$^2$, Symbolic Computation and Satisfiability Checking \cite{AAB+16a}, since these prize mathematical correctness and seek to understand underlying functional relationships.  However, as most developers would acknowledge, our software usually comes with a range of choices which, while having no effect on the correctness of the end result, could have a great effect on the resources required to find it.  These choices range from the low level (in what order to perform a search that may terminate early) to the high (which of a set of competing exact algorithms to use for this problem instance).  In making such choices we may be faced with decisions where relationships are not fully understood, but are not the key object of study.  

In practice such choices may be made by man-made heuristics based on some experimentation (e.g. \cite{DSS04}) or \emph{magic constants} where crossing a single threshold changes system behaviour \cite{Carette2004}.  It is likely that many of these decisions could be improved by allowing learning algorithms to analyse the data.  The broad topic of this paper is ML for algorithm choices where the input is a set of polynomials, which encompasses a variety of tools in computer algebra systems and the SMT theory of [Quantifier Free] Non-Linear Real Arithmetic, [QF]NRA.  

There has been little research on the use of ML in computer algebra: only \cite{HEWDPB14} \cite{HEWBDP19} \cite{EF19} on the topic of CAD variable ordering choice; \cite{HEDP16}, \cite{HEWBDP19} on the question of whether to precondition CAD with Groebner Bases; and \cite{KIMA16} on deciding the order of sub-formulae solving for a QE procedure.
Within SMT there has been significant work on the Boolean logic side e.g. the portfolio SAT solver \textsc{SATZilla} \cite{XHHL08} and \textsc{MapleSAT} \cite{LHPCG17} which views solver branching as an optimisation problem.  However there is little work on the use of ML to choose or optimise theory solvers.  We note that other fields of mathematical software are ahead in the use of ML, most notably the automated reasoning community (see e.g. \cite{Urban2007},  \cite{KBKU13}, \cite{BHP14}, or the brief survey in \cite{England2018}).

\subsection{Difficulties with ML for problems in NRA}
\label{SUBSEC:Difficulties}

There are difficulties in applying standard ML techniques to problems in NRA.  One is the lack of sufficiently large datasets, which is addressed only partially by the SMT-LIB.  The experiment in \cite{HEDP16} found that the [QF]NRA sections of the SMT-LIB too uniform, and had to resort to random generated examples 
(although the state of benchmarking in computer algebra is far worse \cite{ED16b}).  There have been improvements since then, with the benchmarks increasing both in number and diversity of underlying application.  For example, there are now problems arising from biology \cite{BDEEGGHKRSW17}, \cite{EEGRSW17}   and economics \cite{MBDET18}, \cite{MDE18}.

Another difficulty is the identification of suitable features from the input with which to train the ML models.  There are some obvious candidates concerning the size and degrees of polynomials, and the distribution of variables.  However, this provides a starting set (i.e. before any feature selection takes place) that is small in comparison to other machine learning applications. The main focus of this paper is to introduce a method to automatically (and cheaply) generate further features for ML from polynomial systems.

\subsection{Contribution and plan}
\label{SUBSEC:Plan}

Our main contributions are the new feature generation approach described in Section \ref{SEC:FeatureGen} and the validation of its use in the experiments described in Sections \ref{SEC:MLMethodology}$-$\ref{SEC:Results}.  The experiments are for the choice of variable ordering for cylindrical algebraic decomposition, a topic whose background we first present in Section \ref{SEC:Background}, but we emphasise that the techniques may be applicable more broadly.%  We finish with some final thoughts in Section \ref{SEC:Final}.

\section{Background on variable ordering for CAD}
\label{SEC:Background}

\subsection{Cylindrical algebraic decomposition}
\label{SUBSECT:CAD}

A \emph{Cylindrical Algebraic Decomposition} (CAD) is a \emph{decomposition} of ordered $\mathbb{R}^n$ space into cells arranged \emph{cylindrically}: the projections of any pair of cells with respect to the variable ordering are either equal or disjoint.  The projections form an induced CAD of the lower dimensional space.  The cells are (semi)-algebraic meaning each can be described with a finite sequence of polynomial constraints.  

A CAD is produced to be \emph{truth-invariant} for a logical formula (so the formula is either true or false on each cell).  Such a decomposition can then be used to perform Quantifier Elimination (QE) over the reals, i.e. given a quantified Tarski formula find an equivalent quantifier free formula over the reals.  For example, QE would transform $\exists x, ax^2 + b x + c = 0 \land a \neq 0$ to the equivalent unquantified statement $b^2 - 4ac \geq 0$.  A CAD over the $(x,a,b,c)$-space could be used to ascertain this, so long as the variable ordering ensured that there was an induced CAD of $(a,b,c)$-space.  We test one sample point per cell and construct a quantifier free formula from the relevant semi-algebraic cell descriptions.

CAD was introduced by Collins in 1975 \cite{Collins1975} and works relative to a set of polynomials.  Collins' CAD produces a decomposition so that each polynomial has constant sign on each cell (thus truth-invariant for any formula built with those polynomials).  The algorithm first projects the polynomials into smaller and smaller dimensions; and then uses these to lift $-$ to incrementally build decompositions of larger and larger spaces according to the polynomials at that level.  For further details on CAD see for example the collection \cite{CJ98}.

QE has numerous applications throughout science and engineering \cite{Sturm2006}.  Our work also speeds up independent applications of CAD, such as reasoning with multi-valued functions \cite{DBEW12} or motion planning \cite{WDEB13}. 

\subsection{Variable ordering}
\label{SUBSEC:VarOrd}

The definition of cylindricity and both stages of the algorithm are relative to an ordering of the variables.  
For example, given polynomials in variables ordered as $x_n \succ x_{n-1} \succ \dots, \succ x_2 \succ x_1$ we first project away $x_n$ and so on until we are left with polynomials univariate in $x_1$.  
We then start lifting by decomposing the $x_1-$axis, and then the $(x_1, x_2)-$plane and so so on.  The cylindricity condition refers to projections of cells in $\mathbb{R}^n$ onto a space $(x_1, \dots, x_m$) where $m<n$. There have been numerous advances to CAD since its inception: new projection schemes \cite{McCallum1998}, \cite{MPP19}; partial construction \cite{CH91}, \cite{WBDE14}; symbolic-numeric lifting \cite{Strzebonski2006}, \cite{IYAY09}; adapting to the Boolean structure \cite{BDEMW16}, \cite{EBD15}; and adaptations for SMT \cite{JdM12}, \cite{Brown2013}.  However, in all cases, the need for a fixed variable ordering remains.

Depending on the application, the variable ordering may be determined, constrained, or  free. QE, requires that quantified variables are eliminated first and that variables are eliminated in the order in which they are quantified.  However, variables in blocks of the same quantifier (and the free variables) can be swapped, so there is partial freedom. Of course, in the SMT context there is only a single existentially quantified block and so there is a free choice of ordering. So the discriminant in the example above could have been found with any CAD which eliminates $x$ first.  A CAD for the quadratic polynomial under ordering $a \prec b \prec c$ has only 27 cells, but needs 115 for the reverse ordering.

Since we can switch the order of quantified variables in a statement when the quantifier is the same, we also have some choice on the ordering of quantified variables.  For example, a QE problem of the form $\exists x \exists y \forall a \, \phi(x, y, a)$ could be solved by a CAD under either ordering $x \succ y \succ a$ or ordering $y \succ x \succ a$.

The choice of variable ordering can have a great effect on the time and memory use of CAD, and the number of cells in the output.  Further, Brown and Davenport presented a class of problems in which one variable ordering gave output of double exponential complexity in the number of variables and another output of a constant size \cite{BD07}.  

\subsection{Prior work on choosing the variable ordering}

Heuristics have been developed to choose a variable ordering, with Dolzmann \textit{et al.} \cite{DSS04} giving the best known study.  After analysing a variety of metrics they proposed a heuristic, \texttt{sotd}, which constructs the full set of projection polynomials for each permitted ordering and selects the ordering whose corresponding set has the lowest \textbf{s}um \textbf{o}f \textbf{t}otal \textbf{d}egrees for each of the monomials in each of the polynomials.   The second author demonstrated examples for which that heuristic could be misled in \cite{BDEW13}; and then later showed that tailoring to an implementation could improve performance \cite{EBDW14}.  These heuristics all involved potentially costly projection operations on the input polynomials.

In \cite{HEWDPB14} the second author of the present paper collaborated to use a support vector machine to choose which of three human made heuristics to believe when picking the variable ordering, based only on simple features of the input polynomials.  The experiments identified substantial subclasses on which each of the three heuristics made the best decision, and demonstrated that the machine learned choice did significantly better than any one heuristic overall.  This work was picked up again in \cite{EF19} by the present authors, where ML was used to predict directly the variable ordering for CAD, leading to the shortest computing time, with experiments conducted for four different ML models.  

Both \cite{HEWDPB14} and \cite{EF19} used a set of $11$ human identified features.  These did lead to good performance of the models, with ML outperforming the prior human created heuristics, but a starting set of 11 features is relatively small for ML and so we hypothesise that identifying more would improve the results.

\section{Generating new features algorithmically}
\label{SEC:FeatureGen}

\subsection{Existing features for sets of polynomials}
\label{SUBSEC:OriginalFeatures}

An early heuristic for the choice of CAD variable ordering is that of Brown \cite{Brown2004}, which chooses a variable ordering according to the following criteria, starting with the first and breaking ties with successive ones.
\begin{enumerate}[(1)]
\item Eliminate a variable first if it appears with the lowest overall degree in the input.
\item For each variable calculate the maximum total degree for the set of terms in the input in which it occurs.  Eliminate first the variable for which this is lowest.
\item Eliminate a variable first if there is a smaller number of terms in the input  which contain the variable.
\end{enumerate}
Despite being computationally cheaper than the \texttt{sotd} heuristic (because the latter performs projections before measuring degrees) experiments in \cite{HEWDPB14} suggested this simpler measure actually performs slightly better, although the key message from those experiments is that there were substantial subsets of problems for which each heuristic made a better choice than the others.

The Brown heuristic inspired almost all the features used by the authors of \cite{HEWDPB14}, \cite{EF19} to perform ML for CAD variable ordering, with the full set of 11 features listed in Table \ref{tab1} (column 3 will be explained later).

\begin{table}[b]
	\caption{Features used by ML in \cite{HEWDPB14} to choose the ordering of 3 variables for CAD.}\label{tab1}
	\centering
	\begin{tabular}{|c|clc|cl|}
		\hline
		\textbf{\quad \# \quad } & \quad & \textbf{Description} & \, & \quad & $f_v$\\
		\textbf{} &  & & & & \\		
		\hline
		1 & &Number of polynomials & & & $P$\\
		2 & &Maximum total degree of polynomials & & & $\max_{m,p} \left(\sum_v d_v^{m,p}\right)$ \\
		3 & &Maximum degree of $ x_1 $ among all polynomials & & & $\max_{m,p} d_1^{m,p}$ \\
		4 & &Maximum degree of $ x_2 $ among all polynomials & & & $\max_{m,p} d_2^{m,p}$ \\
		5 & &Maximum degree of $ x_3 $ among all polynomials & & & $\max_{m,p} d_3^{m,p}$\\
		6 & &Proportion of $ x_1 $ occurring in polynomials  & & & $\text{av}_p \left(      \text{sgn}\left(\sum_{m} d_1^{m,p}\right)\right)$\\
		7 & &Proportion of $ x_2 $ occurring in polynomials & & & $\text{av}_p\left( \text{sgn}\left(\sum_{m} d_2^{m,p}\right)\right)$\\
		8 & &Proportion of $ x_3 $ occurring in polynomials & & & $\text{av}_p\left(     \text{sgn}\left(\sum_{m} d_3^{m,p}\right)\right)$ \\
		9 & &Proportion of $ x_1 $ occurring in monomials & & & $\text{av}_{m,p}\left(\text{sgn}\left(d_1^{m,p}\right)\right)$\\
		10 & &Proportion of $ x_2 $ occurring in monomials & & & $\text{av}_{m,p}\left(\text{sgn}\left(d_2^{m,p}\right)\right)$\\
		11 & &Proportion of $ x_3 $ occurring in monomials & & & $\text{av}_{m,p}\left(\text{sgn}\left(d_3^{m,p}\right)\right)$\\
		\hline
	\end{tabular}
\end{table}

\subsection{A new framework for generating polynomial features}
\label{SUBSEC:Framework}

Our new feature generation procedure is based on the observation that all the measurements taken by the Brown heauristic, and all those features used in \cite{HEWDPB14}, \cite{EF19} can be formalised mathematically using a small number of functions. For simplicity, the following discussion will be restricted to polynomials of $3$ variables as these were used in the following experiments, but everything generalises in an obvious way to $n$ variables. 

Let a problem instance $\boldsymbol{Pr}$ be defined by a set of $P$ polynomials
\begin{equation}
\boldsymbol{Pr}=\{\mathcal{P}_p \, \vert \, p=1,\dots,P\}.
\end{equation}
This is the case for producing a sign-invariant CAD.  Of course, any problem instance consisting of a logical formula whose atoms are polynomial sign conditions can also have such a set extracted. 

In the following we define the notation for polynomial variables and coefficients that will be used throughout the manuscript. Each polynomial with index $p$, for $p=1,\dots,P$ contains a different number of monomials, which will be labelled with index $m$, where $m=1,\dots,M_p$ and $M_p$ denotes the number of monomials in polynomial $p$.  We note that these are just labels and are not setting an ordering themselves. The degrees corresponding to each of the variables $x_1, x_2, x_3$ are a function of $m$ and $p$. These need to be explicitly labelled in order to allow a rigorous definition of our proposed procedure of feature generation.

We next write each polynomial as
\begin{equation}
\mathcal{P}_p=\sum_{m=1}^{M_p} c^{m,p}\cdot x_1^{d_1^{m,p}} x_2^{d_2^{m,p}} x_3^{d_3^{m,p}}, p=1,\dots,P.
\end{equation}
Here, $x_v$ represents the polynomial variables ($v=1,2,3$).
Thus for each monomial in each polynomial there is a tuple $(m,p)$ of positive integers that label it.  Then in turn we denote by $d_v^{m,p}$ the degree of variable $x_v$ in that monomial, and by $c^{m,p}$ the constant coefficient, i.e., tuple superscripts are giving a label for a monomial in a problem. The original indices are simply a labelling and not an ordering of the variables $x_1, x_2, x_3.$ 

Therefore, any one of our problem instances $\boldsymbol{Pr}$ is uniquely represented by a set of sets 
\begin{equation}
\mathbb{S}_{\boldsymbol{Pr}}
= \big\lbrace \big\lbrace \left[ 
c^{m,p},(d_1^{m,p},d_2^{m,p},d_3^{m,p}) 
\right] \, \vert \,  m=1,\dots,M_p
\big\rbrace \, \vert \, p=1,\dots,P \big\rbrace.
\end{equation}

Observe now that each of Brown's measures can be formalised as a vector of features for choosing a variable as follows.
\begin{enumerate}[(1)]
\item Overall degree in the input of a variable:
$\max_{m,p} d_v^{m,p}$.
\item Maximum total degree of those terms in the input in which a variable occurs: \\
$\max_{m,p} \text{sgn}(d_v^{m,p})\cdot(d_1^{m,p}+d_2^{m,p}+d_3^{m,p})$
\item Number of terms in the input which contain the variable: 
$\sum_{m,p} \text{sgn}(d_v^{m,p})$
\end{enumerate}
In the latter two we use the sign function to discriminate between monomials which contain a variable (sign of degree is positive) and those which do not (sign of degree is zero).  Of course the sign of the degree is never negative.

Define now also the averaging functions 
\begin{align*}
\text{av}_m &\triangleq \frac{1}{M_p}\sum_{m}, \quad
\text{av}_p \triangleq \frac{1}{P}\sum_{p}, \quad
\text{av}_{m,p} \triangleq \frac{1}{P}\sum_{p}\frac{1}{M_p}\sum_{m}.
\end{align*}
Then the features in Table \ref{tab1} can be formalised similarly to Brown's metrics, as shown in the third column of Table \ref{tab1}.

We can place all of these scalars into a single framework: 
\begin{equation}
f\left( \boldsymbol{Pr} \right)=(g_4\circ g_3\circ g_2\circ g_1\circ h^{m,p})\left( \boldsymbol{Pr} \right),
\end{equation}
where 
\[
h^{m,p}\left( \boldsymbol{Pr} \right) \in \big\{ d_v^{m,p}, \, 
\text{sgn} \left(d_v^{m,p}\right) \cdot \left( \textstyle \sum_{v'} d_{v'}^{m,p}\right) \, \vert \, v=1,2,3 \big\}
\]
and $g_1, g_2, g_3$, and $g_4$ are all taken from the set
\[
\big\{ \textstyle \max_p, \max_m, \max_{m,p}, \max_0,\sum_p, \sum_m, \sum_{m,p}, \sum_0, \text{av}_p, \text{av}_m,\text{av}_{m,p}, \text{av}_0, \text{sgn}, \text{sgn}_0 \big\}.
\]
In the above set $\max_0$, $\sum_0$, $\text{av}_0$ and $\text{sgn}_0$ are all equal to the identity function.

For example, let $\boldsymbol{Pr}=\{x_1^2x_2-x_3, x_1  x_2^4 x_3^2+ x_1 x_3\}$. If $m=1, p=2$, then $h^{1,2}\left( \boldsymbol{Pr} \right) \in \big\{ d_v^{1,2}, \, 
\text{sgn} \left(d_v^{1,2}\right) \cdot \left( \textstyle \sum_{v'} d_{v'}^{1,2}\right) \, \vert \, v=1,2,3 \big\}= \big\{ 1,1\cdot 7, 4, 4\cdot 7,2,2\cdot 7 \big\}$.

\subsection{Generating additional features}
\label{SUBSEC:NewFeatures}

We will thus consider deriving all of the other features which fall into this framework, but to do so we must first impose a number of rules.
\begin{enumerate}
%\item Two functions $g_{i_1}, g_{i_2}$ cannot be part of the same category, for the categories the set $\{\max, \sum, $ $\text{av},\text{sgn}\}.$
\item The functions $g_1,g_2,g_3,g_4$ must all belong to distinct categories of function, i.e. one each of $\max$, $\sum$, $\text{av}$, and $\text{sgn}$.
\item Exactly one of the functions $g_1,g_2,g_3,g_4$ is computed over $p$ and exactly one is computed over $m$ (it may be the same one).
\item The computation over $p$ is always performed by a function $g_i$ with an index $i$ greater or equal to that of the function computing over $m$.
\end{enumerate}

\begin{wraptable}{r}{4.2cm}
\caption{Possible distributions of indices to the function classes in feature framework. \label{tab:1728}}

\vspace*{0.1in}

\begin{tabular}{cccc}
max 	& av 	& sum 	& sgn \\
\hline
$p,m$ 	& $0$ 	& $0$ 	& 0\\
$p$		& $m$ 	& $0$ 	& 0\\
$p$ 	& $0$ 	& $m$ 	& 0\\
$0$		& $p,m$	& $0$ 	& 0\\
$0$		& $p$ 	& $m$ 	& 0\\
$0$		& $0$	& $p,m$ & 0\\
$p,m$ 	& $0$ 	& $0$ 	& 1\\
$p$		& $m$ 	& $0$ 	& 1\\
$p$ 	& $0$ 	& $0$ 	& 1\\
$0$		& $p,m$	& $0$ 	& 1\\
$0$		& $p$ 	& $m$ 	& 1\\
$0$		& $0$	& $p,m$ & 1\\
\end{tabular}
\centering
\end{wraptable}

The expression of $f\left( \boldsymbol{Pr} \right)$ can be interpreted as follows. The values $h^{m,p}\left( \boldsymbol{Pr} \right)$ are  functions of variables $m$ and $p$. Each of the functions $g_1, g_2, g_3, g_4$ either leave the function unchanged, or they turn it into a function of fewer variables (first into a function of $p$, and then into a scalar value, representing the ML feature).  

The rules above are justified as follows. Rule $1$ reduces the redundancy in the feature set. Rules $2$ and $3$ guarantee that the feature $f_v\left( \boldsymbol{Pr} \right)$ is well defined and is a scalar number. In particular, Rule $3$ is necessary because the computation over the terms in a polynomial is dependent on their number, which is not the same for all polynomials.

The final set $\{f^{(1)}(\boldsymbol{Pr}),\dots,f^{(N_f)}(\boldsymbol{Pr})\}$ has size $N_f=1728$ for a problem with $3$ variables. 
This number is attained as follows: we have 12 possible distributions of  indexes to the functions $g_1, \dots, g_4$ as shown in Table \ref{tab:1728}; then $4!$ possible orderings of those functions; and 6 possible choices for $h$.  $4! \cdot 6 \cdot 12 = 1728$.

However, many of these features will be identical (e.g. to a different placement of the identify function).  We do not identify these manually now: the task that is trivial for a given dataset, but substantial to do in generality.

%There are $1728$ features generated as above for a problem with $3$ variables. 

\section{Machine learning experiment with the new features}
\label{SEC:MLMethodology}

We now describe a ML experiment to choose the variable ordering for cylcindrical algebraic decomposition.  The methodology here is similar to that in our recent paper \cite{EF19} except for the addition of the extra features from Section \ref{SEC:FeatureGen}.  A more detailed discussion of the methodology can be found in \cite{EF19}.

\subsection{Problem set}
\label{SUBSEC:dataset}

We use the \texttt{nlsat} dataset\footnote{Freely available from \url{http://cs.nyu.edu/~dejan/nonlinear/}} produced to evaluate the work in \cite{JdM12}, thus the problems are all fully existentially quantified.  
Although there are CAD algorithms that reduce what is being computed based on the quantifiers in the input (most notably via Partial CAD \cite{CH91}), the conclusions drawn are likely to be applicable outside of the SAT context.  

We use the $6117$ problems with $3$ variables from this database, so each has a choice of six different variable orderings.  We extracted only the polynomials involved, and randomly divided into two datasets for training ($ 4612 $) and testing ($ 1505 $).  Only the former was used to tune the parameters of the ML models.
% and the testing dataset was unknown to the models during training, and is used to compare the performance of the different ML models, and to compare with the human constructed heuristics. 

\subsection{Software}
\label{SUBSEC:software}

We used the CAD routine \texttt{
CylindricalAlgebraicDecompose} which is part of the \texttt{RegularChains} Library for \textsc{Maple}.   This algorithm builds decompositions first of $n$-dimensional complex space before refining to a CAD of $\mathbb{R}^n$ \cite{CMXY09}, \cite{CM14b}, \cite{BCDEMW14}.  We ran the code in Maple $ 2018 $ but used an updated version of the \texttt{RegularChains} Library (\url{http://www.regularchains.org}).  
Training and evaluation of the ML models was done using the  \texttt{scikit-learn} package \cite{SciKitLearn2011} v0.20.2 for Python 2.7.  The features for ML were extracted using code written in the \texttt{sympy} package 
%\cite{SymPy2017} 
v1.3 for Python 2.7, as was Brown's heuristic.  The \texttt{sotd} heuristic was implemented in \textsc{Maple} as part of the \texttt{ProjectionCAD} package \cite{EWBD14}. 

\subsection{Timings}
\label{SUBSEC:Timing}

CAD construction was timed in a Maple script that was called separately from Python for each CAD (to avoid Maple's caching of results).  The target variable ordering for ML was defined as the one that minimises the computing time for a given problem.  
All CAD function calls included a time limit. For the training dataset an initial time limit of $4$ seconds was used, doubled incrementally if all orderings timed out, until CAD completed for at least one ordering (a target variable ordering could be assigned for all problems using time limits no bigger than $64$ seconds).
The problems in the testing dataset were processed with a larger time limit of $ 128 $ seconds for all orderings (time outs set as $128$s).

\subsection{Feature simplification}
\label{SUBSEC:FeatureSimp}

When computed on a set of problems $\{\boldsymbol{Pr}_1,\dots,\boldsymbol{Pr}_N\}$, some of the features $f^{(i)}$
%,\dots,f^{(N_f)}$ 
turn out to be constant, i.e. 
$
f^{(i)}(\boldsymbol{Pr}_1)=f^{(i)}(\boldsymbol{Pr}_2)=\dots=f^{(i)}(\boldsymbol{Pr}_N).
$
Such features will have no benefit for ML and are removed.  Further, other features may be repetitive, i.e.
$
f^{(i)}(\boldsymbol{Pr}_n)=f^{(j)}(\boldsymbol{Pr}_n),\forall n =1,\dots,N.
$  
This repetition may represent a mathematical equality, or just be the case of the given dataset.  Either way, they are merged into a single feature for the experiment. 
After this step, we are left with $78$ features: so while a large majority were redundant, we still have seven times those available in \cite{HEWDPB14}, \cite{EF19}.

\subsection{Feature selection}
\label{SUBSEC:FeatureSelect}

Feature selection was performed with the training dataset to see if any features were redundant for the ML. We chose the Analysis of Variance (ANOVA) F-value to determine the importance of each feature for the classification task. Other choices we considered were unsuitable for our problem, e.g. the mutual information based selection requires very large amounts of data. 

The training dataset consists of $N=6117$ problems with $3$ variables, and each problem is assigned a target ordering, or class $c=1,\dots,C$, where $C=6$.  Let $\boldsymbol{Pr}_{c,n}$ denote problem number $n$ from the training dataset that is assigned class number $c$, $c=1,\dots,C$ and $n=1,\dots,N_c$, where $N_c$ denotes the number of problems that are assigned class $c.$ Thus $\sum_{c=1}^C N_c=N.$

The F-value for feature number $i$ is computed as follows \cite{MM90}.
\begin{equation}\label{eq:fvalue}
F_i=\frac{\frac{1}{C-1}\sum_{c=1}^C N_c\left(\bar{f}_c^{(i)}-\bar{f}^{(i)}\right)^2}{\frac{1}{N-C}\sum_{c=1}^C \sum_{n=1}^{N_c} \left( f^{(i)}(\boldsymbol{Pr}_{c,n})-\bar{f}_c^{(i)} \right)^2},
\end{equation}
where $\bar{f}_c^{(i)}$ is the sample mean in class $c$, 
%\[
%\bar{f}_c^{(i)}=\frac{1}{N_c}\sum_{n=1}^{N_c} f^{(i)}\left( %\boldsymbol{Pr}_{c,n} \right)
%\]
and $\bar{f}^{(i)}$ the overall mean of the data:
%\[
%\bar{f}^{(i)}=\frac{1}{N} \sum_{c=1}^C \sum_{n=1}^{N_c} f^{(i)}(\boldsymbol{Pr}_{c,n}).
%\]
\begin{align*}
\bar{f}_c^{(i)} &=
\frac{1}{N_c}\sum_{n=1}^{N_c} f^{(i)}\left( \boldsymbol{Pr}_{c,n} \right),
\qquad
\bar{f}^{(i)} =
\frac{1}{N} \sum_{c=1}^C \sum_{n=1}^{N_c} f^{(i)}(\boldsymbol{Pr}_{c,n}).
\end{align*}
The numerator in \eqref{eq:fvalue} represents the \textit{between-class variability} or \textit{explained variance} and the denominator the \textit{within-class variability} or \textit{unexplained variance}. 

%The features with small $F_i$ are less relevant for the classification task, and  therefore will be removed first from the dataset.

Of the 78 features the three with the highest F-values were the following
\begin{align*}
f_{65}\left(\boldsymbol{Pr}\right)&=
\textstyle
\max_0 \text{av}_{m,p} \sum_0 \text{sign} \left(d_2^{m,p}\right)= \frac{1}{P}\sum_{p} \frac{1}{M_p} \sum_{m} \text{sign}\left(d_2^{m,p}\right)
\\
f_{46}\left(\boldsymbol{Pr}\right)&=
\textstyle
\max_0 \sum_{p} \text{av}_{m} \text{sign}\left(d_2^{m,p}\right) \cdot \left( \sum_{v'} d_{v'}^{m,p}\right)
\\
&= \textstyle \sum_{p} \frac{1}{M_p} \sum_{m} \text{sign}\left(d_2^{m,p}\right) \cdot \left( \sum_{v'} d_{v'}^{m,p}\right)
\\
f_{76}\left(\boldsymbol{Pr}\right)&=
\textstyle
\text{av}_{0} \sum_{p} \max_m   \text{sign}\left(d_2^{m,p}\right) \cdot \left( \sum_{v'} d_{v'}^{m,p}\right)\\
&= \textstyle \sum_{p} \max_m \text{sign}\left(d_2^{m,p}\right) \cdot \left( \sum_{v'} d_{v'}^{m,p}\right)
\end{align*}
The new features may be translated back into natural language. For example, feature $65$ is the proportion of monomials containing variable $x_2$, averaged across all polynomials; feature $46$ the sum of the degrees of the variables in all monomials containing variable $x_2$, averaged across all monomials and summed across all polynomials;and feature $76$ the maximum sum of the degrees of the variables in all monomials containing variable $x_2$, summed across all polynomials.

Feature selection did not suggest to remove any features (they all contributed meaningful information), so we proceed with our experiment using all 78.

\subsection{ML models}

Four of the most commonly used deterministic ML models were tuned on the training data (for details on the methods see e.g. the textbook \cite{Bishop2006}).  
%We also give an overview of each in \cite{EF19}.
\begin{itemize}
\item The K$-$Nearest Neighbours (KNN) classifier \cite[\S 2.5]{Bishop2006}.
\item The Multi-Layer Perceptron (MLP) classifier \cite[\S 2.5]{Bishop2006}.
\item The Decision Tree (DT) classifier \cite[\S 14.4]{Bishop2006}.
\item The Support Vector Machine (SVM) classifier with RBF kernel \cite[\S 6.3]{Bishop2006}. 
\end{itemize}

Each model was trained using grid search $5$-fold cross-validation, i.e. the set was randomly divided into $5$ and each possible combination of $4$ parts was used to tune the model parameters, leaving the last part for fitting the hyperparameters with cross-validation, by optimising the average F-score.
Grid searches were performed for an initially large range for each hyperparameter; then gradually decreased to home into optimal values. This lasted from a few seconds for simpler models like KNN to a few minutes for more complex models like MLP. The optimal hyperparameters selected during cross-validation are in Table \ref{tab3}.

%\begin{table}[t]
%	\caption{The hyperparameters of the ML models optimised with $ 5 $-fold cross-validation on the training dataset.}\label{tab3}
%
%	\begin{center}
%		Decision Tree\\
%		\vspace{0.1cm}
%		\begin{tabular}{|c|c|}
%			\hline
%			\textbf{Hyperparameter} & \textbf{Value} \\
%			\hline
%			Criterion & Gini impurity\\			
%			Maximum tree depth & $ 17 $\\
%			\hline
%		\end{tabular}\\
%		\vspace{0.5cm}
%		K-Nearest Neighbours\\
%		\vspace{0.1cm}
%		\begin{tabular}{|c|c|}
%			\hline
%			\textbf{Hyperparameter} & \textbf{Value} \\
%			\hline
%			Train instances weighting & Inversely proportional to distance\\			
%			Algorithm & Ball Tree\\
%			\hline
%		\end{tabular}\\	
%		\vspace{0.5cm}
%		Support Vector Machine\\
%		\vspace{0.1cm}
%		\begin{tabular}{|c|c|}
%			\hline
%			\textbf{Hyperparameter} & \textbf{Value} \\
%			\hline
%			Regularization parameter $ C $ & $ 316 $\\			
%			Kernel  & Radial basis function\\
%			$ \gamma $ & $ 0.08 $\\
%			Tolerance for stopping criterion & $ 0.0316 $\\
%			\hline
%		\end{tabular}\\	
%		\vspace{0.5cm}
%		Multi-Layer Perceptron\\
%		\vspace{0.1cm}
%		\begin{tabular}{|c|c|}
%			\hline
%			\textbf{Hyperparameter} & \textbf{Value} \\
%			\hline
%			Hidden layer size & $ 18 $\\			
%			Activation function  & Hyperbolic tangent\\
%			Algorithm & Quasi-Newton based optimiser\\
%			Regularization parameter $ \alpha $ & $ 5\cdot10^{-5} $\\
%			\hline
%		\end{tabular}	
%	\end{center}
%\end{table}

\begin{table}[t]
	\caption{The ML hyperparameters used following optimisation on the training dataset.}\label{tab3}
\begin{tabular}{|c|c|c|}
\hline
\textbf{Model} & \textbf{Hyperparameter} & \textbf{Value} \\
\hline
\hline
Decision Tree 	& Criterion & Gini impurity\\			
				& Maximum tree depth & $ 17 $\\
\hline
K-Nearest  & Train instances weighting & Inversely proportional to distance\\			
Neighbours			& Algorithm & Ball Tree\\
\hline
Support Vector  & Regularization parameter $ C $ & $ 316 $\\			
Machine		&	Kernel  & Radial basis function\\
		&	$ \gamma $ & $ 0.08 $\\
		&	Tolerance for stopping criterion & $ 0.0316 $\\
			\hline
Multi-Layer  & Hidden layer size & $ 18 $\\			
Perceptron		&	Activation function  & Hyperbolic tangent\\
	&		Algorithm & Quasi-Newton based optimiser\\
	&		Regularization parameter $ \alpha $ & $ 5\cdot10^{-5} $\\
	\hline
		\end{tabular}	
\end{table}

\subsection{Comparing with human made heuristics}
\label{SUBSEC:Human}
The ML approaches were compared in terms of prediction accuracy and resulting CAD computing time against the two best known human constructed heuristics \cite{Brown2004}, \cite{DSS04} as discussed earlier.
Unlike the ML, these can end up predicting several variable orderings (i.e. when they cannot discriminate).  In practice if this were to happen the heuristic would select one randomly (or perhaps lexicographically), however that final pick is not meaningful.   To accommodate this, for each problem, the prediction accuracy of such a heuristic is judged to be the the percentage of its predicted variable orderings that are also target orderings.  The average of this percentage over all problems in the testing dataset represents the prediction accuracy.  Similarly, the computing time for such methods was assessed as the average computing time over all predicted orderings, and it is this that is summed up for all problems in the testing dataset.

\section{Experimental Results}
\label{SEC:Results}

The results are presented in Table \ref{tab2}.  We compare the four ML models on the percentage of problems where they selected the optimum ordering, and the total computation time (in seconds) for solving all the problems with their chosen orderings.  The first two rows reproduce the results of \cite{EF19} which used only the 11 features from Table \ref{tab1}, while the latter two rows are the results from the new experiment in the present paper which has 78 features. 
We also compare with the two human constructed heuristics and the outcome of a random choice between the 6 orderings (which do not change with the number of features).
We might expect a random choice to be correct one sixth of the time but it is higher as for some problems there were multiple variable orderings with equally fast timings.

We also consider the distribution of the computation times: the differences between the computation time of each method and the minimum computation time, given as a percentage of the minimum time, are depicted in Figure 1.

\begin{table}[b]
	\caption{The comparative performance of DT, KNN, MLP, SVM, and the Brown and sotd heuristics on the testing dataset for the present experiment and the one in \cite{EF19}.}\label{tab2}
	\begin{center}
		\begin{tabular}{|l|l|c|c|c|c|c|c|c|}
			\hline
			 & &  DT & KNN & MLP & SVM & \texttt{Brown} & \texttt{sotd} & rand\\
			\hline
	From \cite{EF19}	&	\textbf{Accuracy}  &  $ 62.6 \% $ & $ 63.3 \% $ & $ 61.6 \% $ & $ 58.8 \% $ & $ 51 \% $ & $ 49.5 \% $ & $ 22.7 \% $\\
	(11 Features)	&	\textbf{Time (s)} & $ 9\,994 $ & $ 10\,105 $ & $ 9\,822 $ & $ 10\,725 $ & $ 10\,951 $ & $ 11\,938 $ &  $ 30\,235 $ \\				
			\hline
	New Experiment	&	\textbf{Accuracy}  &  $ 65.2 \% $ & $ 66.3 \% $ & $ 67 \% $ & $ 65 \% $ &  &  & \\
	(78 Features)	&	\textbf{Time (s)} & $ 9\,603 $ & $ 9\,178 $ & $ 9\,399 $ & $ 9\,487 $ &   &  &   \\				
			\hline
		\end{tabular}
	\end{center}
\end{table}

\begin{figure}[t]
\centering
	\includegraphics[width=0.85\textwidth]{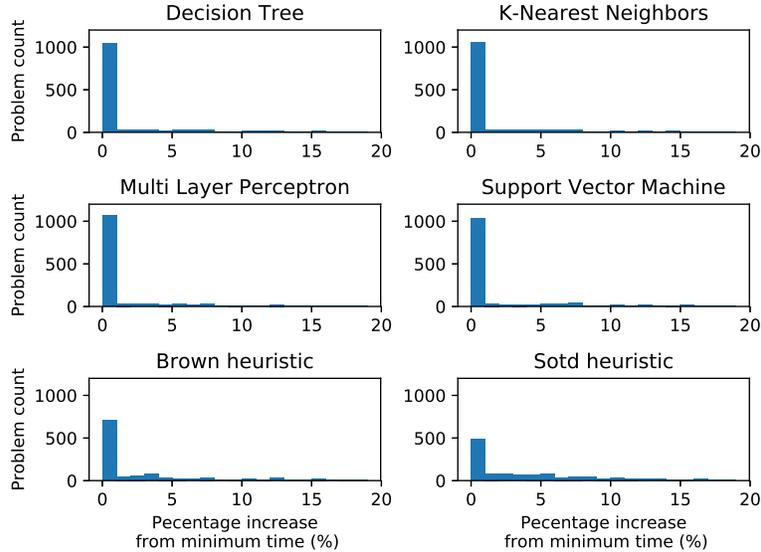}
	\caption{The histograms of the percentage increase in computation time relative to the minimum computation time for each method, calculated for a bin size of 1\%.} \label{fig1}
\end{figure}

\subsection{Range of possible outcomes}

The minimum total computing time, achieved if we select an optimal ordering for every problem, is $8\,623$s. Choosing at random would take $ 30\,235 $s, almost 4 times as much. The maximum time, if we selected the worst ordering for every problem, is $ 64\,534 $s. The K-Nearest Neighbours model achieved the shortest time of our models and heuristics, with $9\,178$s, only $ 6\% $ more than the minimal possible. %So there are clearly great time savings to be made by taking this choice into account.

\subsection{Human-made heuristics}

Since they are not affected by the new feature framework of the present paper the findings on the human made heuristics are the same as in \cite{EF19}. Of the two human-made heuristics, \texttt{Brown} performed the best, 
%, as it did in \cite{HEWDPB14}.  As was noted there this is 
surprising since the \texttt{sotd} heuristic has access to additional information (not just the input polynomials but also their projections). Obtaining an ordering for a problem instance with \texttt{sotd} hence takes longer than for Brown or any ML model $-$ generating an ordering with \texttt{sotd} for all problems in the testing dataset took over $ 30 $min.
Using \texttt{Brown} we can solve all problems in 10,951s, 27\% more than the minimum.  While \texttt{sotd} is only 0.7\% less accurate than \texttt{Brown} in identifying the best ordering, it is much slower at $11\,938$s or 38\% more than the minimum.  So, while \texttt{Brown} is not much better at identifying the best, it is much better at discarding the worst!

\subsection{ML choices}

The results show that all ML approaches outperform the human constructed heuristics in terms of both accuracy and timings. Moreover, the results show that the new algorithm for generating features leads to a clear improvement in ML performance compared to using only a small number of human generated features in \cite{EF19}.  For all four modules both accuracy has increased and computation time decreased. 
%Specifically, the best performance for ML in \cite{EF19} was a computing time of $9\,822$ seconds using the MLP model, and an accuracy of $63.3\% $ with the KNN model. Using the additional $67$ features generated in this study led to a computing time shorter by $622 $ seconds, and an accuracy $3\% $ higher, although it is now the MLP with highest accuracy and the KNN the lowest time.
The best achieved time was 14\% above the minimum using the original 11 features but now only 6\% above with the new features.  

The computing time for all the methods lies between the best ($8\,623$s) and the worst ($64\,534$s). Therefore, if we scale this time to $[0,100]$ so that the shortest time corresponds to $0$ and the slowest to $100$, then the best human-made heuristic (Brown) lies at $4.16$, and the best ML method (KNN) lies at $0.99$. So using ML allows us to be $4$ times closer to the minimum possible computing time.

Figure \ref{fig1} shows that the human-made heuristics result in computing times that are often significantly larger than $ 1\% $ of the corresponding minimum time for each problem. The ML methods, on the other hand, all result in over $ 1000 $ problems ($\sim 75\% $ of the testing dataset) within $ 1\% $ of the minimum time.

\section{Final Thoughts}
\label{SEC:Final}

In this experiment the MLP and KNN models offered the best performance, and a clear advance on the prior state of the art.  
But we acknowledge that there is much more to do and emphasise that these are only the initial findings of the project and we need to see if the findings are replicated.  Planned extensions include:  expanding the dataset to problems with more variables and quantifier structure; trying different feature selection techniques, and seeing if classifiers trained for the \textsc{Maple} CAD may be applied to other implementations.

Our main result is that a great many more features can be obtained trivially from the input (i.e. without any projection operations) than previously thought, and that these are relevant and lead to better ML choices.  Some of these are easy to express in natural language, such as the number of polynomials containing a certain variable, but others do not have an obvious interpretation. This is important because something that is hard to describe in natural language is unlikely to be suggested by a human as a feature, which illustrates the benefit of our framework.  
This contribution to feature extraction for algebraic problems should be more widely applicable than the CAD variable ordering decision.

\subsubsection*{Acknowledgements}  
This work is supported by EPSRC Project EP/R019622/1: \emph{Embedding Machine Learning within Quantifier Elimination Procedures}.

%
% ---- Bibliography ----
%
% BibTeX users should specify bibliography style 'splncs04'.
% References will then be sorted and formatted in the correct style.
%
% \bibliographystyle{splncs04}
% \bibliography{mybibliography}
%

\bibliographystyle{splncs04}
%\bibliography{CAD}

\begin{thebibliography}{10}
\providecommand{\url}[1]{\texttt{#1}}
\providecommand{\urlprefix}{URL }
\providecommand{\doi}[1]{https://doi.org/#1}

\bibitem{AAB+16a}
{\'A}brah{\'a}m, E., et al.:
  $\mathsf{SC}^2$: Satisfiability checking meets symbolic computation. In:
  Proc. CICM '16, LNCS 9791, pp. 28--43. Springer
  (2016)
  %, \url{https://doi.org/10.1007/978-3-319-42547-4_3}

\bibitem{Bishop2006}
Bishop, C.: Pattern Recognition and Machine Learning. Springer (2006)

\bibitem{BCDEMW14}
Bradford, R., Chen, C., Davenport, J., England, M., {Moreno~Maza}, M., Wilson,
  D.: Truth table invariant cylindrical algebraic decomposition by regular
  chains. In: Proc. CASC '14, LNCS 8660, pp. 44--58. Springer  (2014)
  %, \url{http://dx.doi.org/10.1007/978-3-319-10515-4_4}

\bibitem{BDEEGGHKRSW17}
Bradford, R., et al.: A case study on
  the parametric occurrence of multiple steady states. In: Proc. ISSAC '17, pp.
  45--52. ACM (2017)
  %, \url{https://doi.org/10.1145/3087604.3087622}

\bibitem{BDEMW16}
Bradford, R., Davenport, J., England, M., McCallum, S., Wilson, D.: Truth table
  invariant cylindrical algebraic decomposition. J. Symb.
  Comp.  \textbf{76},  1--35 (2016)
  %,  \url{http://dx.doi.org/10.1016/j.jsc.2015.11.002}

\bibitem{BDEW13}
Bradford, R., Davenport, J., England, M., Wilson, D.: Optimising problem
  formulations for cylindrical algebraic decomposition. In: Intelligent
  Computer Mathematics, LNCS 7961, pp.
  19--34. Springer Berlin Heidelberg (2013)
  %,  \url{http://dx.doi.org/10.1007/978-3-642-39320-4_2}

\bibitem{BHP14}
Bridge, J., Holden, S., Paulson, L.: Machine learning for first-order theorem
  proving. Journal of Automated Reasoning  \textbf{53},  141--172 (2014)
  %,  \url{https://doi.org/10.1007/s10817-014-9301-5}

\bibitem{Brown2004}
Brown, C.: Tutorial: {C}ylindrical algebraic decomposition,
  at {ISSAC} '04. 
  \url{http://www.usna.edu/Users/cs/wcbrown/research/ISSAC04/handout.pdf}
  (2004)

\bibitem{Brown2013}
Brown, C.: Constructing a single open cell in a cylindrical algebraic
  decomposition. In: Proc. ISSAC '13, pp. 133--140. ACM
  (2013)
  %, \url{https://doi.org/10.1145/2465506.2465952}

\bibitem{BD07}
Brown, C., Davenport, J.: The complexity of quantifier elimination and
  cylindrical algebraic decomposition. In: Proc. ISSAC '07, pp.
  54--60. ACM (2007)
  %, \url{https://doi.org/10.1145/1277548.1277557}

\bibitem{Carette2004}
Carette, J.: Understanding expression simplification. In: Proc. ISSAC '04,
  pp. 72--79. ACM (2004)
  %, \url{https://doi.org/10.1145/1005285.1005298}

\bibitem{CJ98}
Caviness, B., Johnson, J.: Quantifier Elimination and Cylindrical Algebraic
  Decomposition. Texts \& Monographs in Symbolic Computation, Springer-Verlag
  (1998)
  %, \url{https://doi.org/10.1007/978-3-7091-9459-1}

\bibitem{CM14b}
Chen, C., {Moreno Maza}, M.: An incremental algorithm for computing cylindrical
  algebraic decompositions. In: Computer
  Mathematics, pp. 199---221. Springer Berlin Heidelberg (2014)
  %, \url{https://doi.org/10.1007/978-3-662-43799-5_17}

\bibitem{CMXY09}
Chen, C., {Moreno Maza}, M., Xia, B., Yang, L.: Computing cylindrical algebraic
  decomposition via triangular decomposition. In: Proc. ISSAC '09,
  95--102. ACM (2009)
  %, \url{https://doi.org/10.1145/1576702.1576718}

\bibitem{Collins1975}
Collins, G.: Quantifier elimination for real closed fields by cylindrical
  algebraic decomposition. In: Proc. 2nd GI Conference on Automata
  Theory and Formal Languages. pp. 134--183. Springer-Verlag (reprinted in the
  collection \cite{CJ98}) (1975)
  %, \url{https://doi.org/10.1007/3-540-07407-4_17}

\bibitem{CH91}
Collins, G., Hong, H.: Partial cylindrical algebraic decomposition for
  quantifier elimination. Journal of Symbolic Computation  \textbf{12},
  299--328 (1991)
  %, \url{https://doi.org/10.1016/S0747-7171(08)80152-6}

\bibitem{DBEW12}
Davenport, J., Bradford, R., England, M., Wilson, D.: Program verification in
  the presence of complex numbers, functions with branch cuts etc. In: Proc. SYNASC '12, pp. 83--88. IEEE (2012)
  %,  \url{http://dx.doi.org/10.1109/SYNASC.2012.68}

\bibitem{DSS04}
Dolzmann, A., Seidl, A., Sturm, T.: Efficient projection orders for {CAD}. In:
  Proc. ISSAC '04, pp. 111--118.  ACM (2004)%,
  %\url{https://doi.org/10.1145/1005285.1005303}

\bibitem{England2018}
England, M.: Machine learning for mathematical software. In: Mathematical Software -- Proc. ICMS '18. LNCS 10931, pp. 165--174. Springer
   (2018)%,
  %\url{https://doi.org/10.1007/978-3-319-96418-8_20}

\bibitem{EBD15}
England, M., Bradford, R., Davenport, J.: Improving the use of equational
  constraints in cylindrical algebraic decomposition. In: Proc. ISSAC '15, pp.
  165--172. ACM (2015)%,
  %\url{http://dx.doi.org/10.1145/2755996.2756678}

\bibitem{EBDW14}
England, M., Bradford, R., Davenport, J., Wilson, D.: Choosing a variable
  ordering for truth-table invariant CAD by
  incremental triangular decomposition. In: Proc. ICMS '14, LNCS 8592, pp. 450--457. Springer (2014)%,
  %\url{http://dx.doi.org/10.1007/978-3-662-44199-2_68}

\bibitem{ED16b}
England, M., Davenport, J.: Experience with heuristics, benchmarks \& standards
  for cylindrical algebraic decomposition. In: Proc. $\mathsf{SC}^2$ '16. CEUR-WS \textbf{1804} (2016)%, \url{http://ceur-ws.org/Vol-1804/}

\bibitem{EEGRSW17}
England, M., Errami, H., Grigoriev, D., Radulescu, O., Sturm, T., Weber, A.:
  Symbolic versus numerical computation and visualization of parameter regions
  for multistationarity of biological networks. In: Computer Algebra in Scientific Computing (CASC '17), LNCS 10490, pp. 93--108.
  Springer (2017)%,
  %\url{https://doi.org/10.1007/978-3-319-66320-3_8}

\bibitem{EF19}
England, M., Florescu, D.: Comparing machine learning models to choose the
  variable ordering for cylindrical algebraic decomposition. To appear in Proc.
  CICM '19 (Springer LNCS)  (2019).  Preprint:  \url{https://arxiv.org/abs/1904.11061}

\bibitem{EWBD14}
England, M., Wilson, D., Bradford, R., Davenport, J.: Using the {R}egular
  {C}hains {L}ibrary to build cylindrical algebraic decompositions by
  projecting and lifting. In: Mathematical Software --
  ICMS '14. LNCS 8592, pp. 458--465.
  Springer (2014)%,
  %\url{http://dx.doi.org/10.1007/978-3-662-44199-2_69}

\bibitem{HEDP16}
Huang, Z., England, M., Davenport, J., Paulson, L.: Using machine learning to
  decide when to precondition cylindrical algebraic decomposition with
  {G}roebner bases. In: Proc. SYNASC '16. pp. 45--52. IEEE (2016)%,
  %\url{https://doi.org/10.1109/SYNASC.2016.020}

\bibitem{HEWBDP19}
Huang, Z., England, M., Wilson, D., Bridge, J., Davenport, J., Paulson, L.:
  Using machine learning to improve cylindrical algebraic decomposition.
  Mathematics in Computer Science, Volume to be assigned,  28 pages
  (2019)%, \url{https://doi.org/10.1007/s11786-019-00394-8}

\bibitem{HEWDPB14}
Huang, Z., England, M., Wilson, D., Davenport, J., Paulson, L., Bridge, J.:
  Applying machine learning to the problem of choosing a heuristic to select
  the variable ordering for cylindrical algebraic decomposition. In: Intelligent Computer
  Mathematics,  LNAI 8543, pp.
  92--107. Springer International (2014)%,
  %\url{http://dx.doi.org/10.1007/978-3-319-08434-3_8}

\bibitem{IYAY09}
Iwane, H., Yanami, H., Anai, H., Yokoyama, K.: An effective implementation of a
  symbolic-numeric cylindrical algebraic decomposition for quantifier
  elimination. In: Proc. SNC '09, pp. 55--64. SNC '09 (2009)%,
  %\url{https://doi.org/10.1145/1577190.1577203}

\bibitem{JdM12}
Jovanovic, D., de~Moura, L.: Solving non-linear arithmetic. In: Gramlich, B.,
  Miller, D., Sattler, U. (eds.) Automated Reasoning $-$ Proc. IJCAR '12, LNCS 7364, pp.
  339--354. Springer (2012)%, \url{https://doi.org/10.1007/978-3-642-31365-3_27}

\bibitem{KIMA16}
Kobayashi, M., Iwane, H., Matsuzaki, T., Anai, H.: Efficient subformula orders
  for real quantifier elimination of non-prenex formulas. In: Proc. MACIS '15, LNCS 9582, pp.
  236--251. Springer International Publishing (2016)%,
  %\url{https://doi.org/10.1007/978-3-319-32859-1_21}

\bibitem{KBKU13}
K{\"u}hlwein, D., Blanchette, J., Kaliszyk, C., Urban, J.: {MaSh}: {M}achine
  learning for sledgehammer. In: Interactive Theorem Proving, LNCS 7998, pp. 35--50. Springer Berlin Heidelberg (2013)%,
  %\url{https://doi.org/10.1007/978-3-642-39634-2_6}

\bibitem{LHPCG17}
Liang, J., {Hari Govind}, V., Poupart, P., Czarnecki, K., Ganesh, V.: An
  empirical study of branching heuristics through the lens of global learning
  rate. In: Proc. {SAT} '17, LNCS 10491, pp. 119--135. Springer (2017)%,
 % \url{https://doi.org/10.1007/978-3-319-66263-3_8}

\bibitem{McCallum1998}
McCallum, S.: An improved projection operation for cylindrical algebraic
  decomposition. In: \cite{CJ98}, pp. 242--268. (1998)%,
  %\url{https://doi.org/10.1007/978-3-7091-9459-1_12}

\bibitem{MM90}
Markowski, C.A. and Markowski, E.P.: Conditions for the effectiveness of a preliminary test of variance.
The American Statistician, \textbf{44}:4, 322--326 (1990)

\bibitem{MPP19}
McCallum, S., Parusi\'niski, A., Paunescu, L.: Validity proof of {L}azard's
  method for {CAD} construction. Journal of Symbolic Computation  \textbf{92},
  52--69 (2019)%, \url{https://doi.org/10.1016/j.jsc.2017.12.002}

\bibitem{MBDET18}
Mulligan, C., Bradford, R., Davenport, J., England, M., Tonks, Z.: Non-linear
  real arithmetic benchmarks derived from automated reasoning in economics. In:
  Proc. $\mathsf{SC}^2$ '18, pp.
  48--60. CEUR-WS \textbf{2189} (2018)%,
  %\url{http://ceur-ws.org/Vol-2189/}

\bibitem{MDE18}
Mulligan, C., Davenport, J., England, M.: {T}heory{G}uru: {A} {M}athematica
  package to apply quantifier elimination technology to economics. In:
  Mathematical Software -- Proc. ICMS '18. LNCS 10931, pp.
  369--378. Springer (2018)%,
  %\url{https://doi.org/10.1007/978-3-319-96418-8_44}

\bibitem{SciKitLearn2011}
Pedregosa, F., et al.:
  Scikit-learn: {M}achine learning in {P}ython. Journal of Machine Learning
  Research  \textbf{12},  2825--2830 (2011)%,
  %\url{http://www.jmlr.org/papers/v12/pedregosa11a.html}

\bibitem{Strzebonski2006}
Strzebo\'{n}ski, A.: Cylindrical algebraic decomposition using validated
  numerics. Journal of Symbolic Computation  \textbf{41}(9),  1021--1038
  (2006)%, \url{https://doi.org/10.1016/j.jsc.2006.06.004}

\bibitem{Sturm2006}
Sturm, T.: New domains for applied quantifier elimination. In: Computer Algebra in Scientific Computing,
  LNCS 4194, pp. 295--301. Springer (2006)%, \url{https://doi.org/10.1007/11870814_25}

\bibitem{Urban2007}
Urban, J.: {MaLARea: A} metasystem for automated reasoning in large theories.
  In: Proc. ESARLT '07, CEUR-WS \textbf{257}, p.~14. CEUR-WS (2007)%,
  %\url{http://ceur-ws.org/Vol-257/}

\bibitem{WBDE14}
Wilson, D., Bradford, R., Davenport, J., England, M.: Cylindrical algebraic
  sub-decompositions. Mathematics in Computer Science  \textbf{8},  263--288
  (2014)%, \url{http://dx.doi.org/10.1007/s11786-014-0191-z}

\bibitem{WDEB13}
Wilson, D., Davenport, J., England, M., Bradford, R.: A ``piano movers''
  problem reformulated. In: Proc. SYNASC '13, pp. 53--60. IEEE
  (2013)%, \url{http://dx.doi.org/10.1109/SYNASC.2013.14}

\bibitem{XHHL08}
Xu, L., Hutter, F., Hoos, H., {Leyton-Brown}, K.: {SAT}zilla: {P}ortfolio-based
  algorithm selection for {SAT}. J. Artificial Intelligence Research
  \textbf{32},  565--606 (2008)%, \url{https://doi.org/10.1613/jair.2490}

\end{thebibliography}

\end{document}